# Information-theoretically Secure Regenerating Codes for Distributed Storage

Nihar B. Shah, K. V. Rashmi and P. Vijay Kumar

*Abstract*—Regenerating codes are a class of codes for distributed storage networks that provide reliability and availability of data, and also perform efficient node repair. Another important aspect of a distributed storage network is its security. In this paper, we consider a threat model where an eavesdropper may gain access to the data stored in a subset of the storage nodes, and possibly also, to the data downloaded during repair of some nodes. We provide explicit constructions of regenerating codes that achieve information-theoretic secrecy capacity in this setting.

## I. INTRODUCTION

We consider a distributed storage system consisting of $n$ storage nodes in a network, each having a capacity to store $\alpha$ symbols over a finite field $\mathbb{F}_q$ of size $q$. Data corresponding to $B$ message symbols (the *message*), each drawn uniformly and independently from $\mathbb{F}_q$, is to be dispersed across these $n$ nodes. An end-user (called a data-collector) must be able to *reconstruct* the entire message by downloading the data stored in any subset of $k$ nodes. If data-reconstruction was the only requirement, any $[n, k]$ maximum-distance-separable (MDS) code such as a Reed-Solomon code would suffice.

A second important aspect of a distributed storage system is the handling of node failures. When a storage node fails, it is replaced by a new, empty node. The replacement node is required to obtain the data that was previously stored in the failed node by downloading data from the remaining nodes in the network. A typical means of accomplishing this is to download the entire message from the network, and extract the desired data from it. However, downloading the entire message, when it eventually stores only a fraction $\frac{1}{k}$ of it, is clearly wasteful of the network resources.

Recently, Dimakis et al. [1] introduced a new class of codes called 'regenerating codes' which are efficient with respect to both storage space utilization and the amount of data downloaded for repair (termed *repair-bandwidth*). Regenerating codes permit node repair by downloading $\beta$ symbols from any subset of $d$ ($\geq k$) remaining nodes, and the total repair-bandwidth $d\beta$ is typically much smaller than the message size $B$. In [1] the authors also establish that the parameters involved must necessarily satisfy the bound:

$$B \leq \sum_{i=0}^{k-1} \min\left(\alpha, (d-i)\beta\right) \ . \quad (1)$$

The authors are with the Dept. of ECE, Indian Institute of Science, Bangalore, India. Email: {nihar, rashmikv, vijay}@ece.iisc.ernet.in. P. Vijay Kumar is also an adjunct faculty member of the Electrical Engineering Systems Department at the University of Southern California, Los Angeles, CA 90089-2565.

This work was supported by Infosys Technologies Limited.

It can be deduced (see [1]) that achieving equality in (1), with parameters $B$, $k$ and $d$ fixed, leads to a tradeoff between the storage space $\alpha$ and the repair-bandwidth $d\beta$. In this tradeoff, the case of minimizing $\alpha$ first and then $\beta$ (for fixed $d$) is termed as the minimum storage regenerating (MSR) case, while carrying out the minimization in the reverse order is termed the minimum bandwidth regenerating (MBR) case. More details on the MSR and MBR cases are provided later in the paper. Explicit constructions of MSR and MBR codes achieving this bound can be found in [2], [4]–[6].

The focus of the present paper is on an additional, important aspect of distributed storage systems, namely, security of the data. Nowadays, individuals as well as businesses are increasingly storing their data over untrusted networks. Peer-to-peer storage systems have storage nodes spread out geographically. Such situations make the data prone to prying adversaries that may gain access to the data stored in some of the nodes. An eavesdropper can also gain additional information by listening to the data downloaded during multiple instances of repair of these nodes. It is imperative to prevent such entities from gaining any useful information. The present paper constructs explicit codes which, while satisfying the reconstruction and repair requirements in the distributed storage network, prevents such an eavesdropper from obtaining any information about the original message.

The threat model considered in this paper is as follows. An eavesdropper can gain read-access to the data stored in any set of at-most $\ell$ ($< k$) storage nodes. The eavesdropper may also gain read-access to the data being downloaded during (possibly multiple instances of) repair of some $\ell'$ ($\leq \ell$) of these $\ell$ nodes. Note that the data downloaded by a replacement node during any instance of repair also contains the data that is eventually stored in that node. This is formalized in the following definition.

*Definition 1 ($\{\ell, \ell'\}$ secure distributed storage system):*
Consider a distributed storage system in which an eavesdropper gains access to the data stored in some ($\ell - \ell'$) nodes, and the data stored as well as the data downloaded during repair in some other $\ell'$ nodes. An $\{\ell, \ell'\}$ secure distributed storage system is one in which such an eavesdropper obtains no information about the message.

We assume that the eavesdroppers have unbounded computational power, are passive, non-collusive, and that the underlying code is globally known. As an example of this model, consider a peer-to-peer storage system. The $\ell'$ nodes described above may represent nodes that are in a network belonging to an adversary, thereby allowing the eavesdropper to listen to all the data downloaded as these $\ell'$ nodes undergo (possibly multiple) failures and repairs across time. On the other hand, the ($\ell - \ell'$) nodes may represent the nodes which may be exposed only momentarily, allowing the eavesdropper access to only the data stored.

The problem of providing information-theoretic secrecy in distributed storage systems can be related to the Wiretap Channel II [7] where an eavesdropper, listening to any arbitrary subset (of fixed size) of symbols being transmitted over a noiseless point-to-point channel, obtains essentially no information about the original message. While schemes providing secrecy in a distributed storage system with only the reconstruction requirement would follow from [7], the requirement of addressing node-repair makes the problem harder. Among recent results in the context of distributed storage, the problem of securely disseminating encoded data to the storage nodes is considered in [8], and an analysis of communication and interaction requirements between the nodes is provided. In [9], the authors consider the situation where data is stored over two networks, and an eavesdropper may gain access to any one of these networks. Connections between optimal repair in distributed storage and communication across multiple-access wiretap channels are established in [10].

The system model considered in the present paper is based on the model introduced by Pawar et al. [3]. In [3], the authors consider the case when $\ell' = \ell$ and provide an upper bound on the number of message symbols $B^{(s)}$ that can be stored in the information-theoretically secure system as

$$B^{(s)} \leq \sum_{i=\ell}^{k-1} \min\left(\alpha, (d-i)\beta\right) \ . \quad (2)$$

The bound in (2) can be interpreted in the following intuitive manner. Out of the $k$ nodes to which a data-collector connects, consider the case where the first $\ell$ of these nodes are compromised. Thus, assuming the secrecy goals have been met, these $\ell$ nodes will provide zero information about the message symbols, and only the remaining $(k-\ell)$ nodes in the summation in (1) provide useful information. It can be shown that the bound in (2) is, in fact, an upper bound on the number of message symbols in an information-theoretically secure system for all values of $\ell'$.

In the sequel, notation pertaining to the secure version of the code will frequently be indicated by the superscript $(s)$. For instance, $B^{(s)}$ denotes the number of message symbols in a system with secrecy constraints, and $B$ denotes the number of message symbols in a system without secrecy constraints (i.e., when $\ell = \ell' = 0$). Note that the difference $B - B^{(s)}$ is the price paid for the additional secrecy constraint.

In [3], the authors also show that the MBR code presented in [4] for the parameters $[n, k, d = n-1]$ can be made information-theoretically secure by making use of a nested MDS code in the construction.

In the present paper, we provide explicit constructions for information-theoretically secure MBR and MSR codes for:

1) MBR, all parameters $[n, k, d]$, and
2) MSR, all parameters $[n, k, d \geq 2k-2]$ .

Each of the constructions presented is $\{\ell, \ell'\}$ information-theoretically secure, for all values of $\ell$ and $\ell'$. The secure MBR code presented is optimal for all $\{\ell, \ell'\}$, and the secure MSR code presented is optimal for all values of $\ell$ when $\ell' = 0$. Thus this also establishes the secrecy capacity of such a system for each of these parameter values. It is unknown at present as to whether or not the MSR code presented here is optimal for $\ell' \geq 1$.

The secure codes provided in the present paper are based on our previous work [2], where we construct explicit regenerating codes for the parameters listed above. The codes in [2] are based on a new Product-Matrix (PM) framework. We will call the MBR and MSR codes of [2] as the PM-MBR and PM-MSR codes respectively, and the corresponding secure versions constructed in the present paper as the secure PM-MSR and the secure PM-MBR codes respectively.

While all other regenerating codes in the literature require the number of nodes $n$ to be equal to $d+1$, the PM codes [2] do not pose any such constraint. Thus the PM codes are well suited for distributed storage systems where the number of nodes $n$ may vary in time, or where the connectivity $d$ required for repair may be low. These codes are also linear, i.e., each symbol in the system is a linear combination of the message symbols. As we shall subsequently see, the PM framework possesses two additional attributes that makes it more attractive for constructing secure codes: (a) exact-repair, and (b) data downloaded by a node for repair is independent of the set of $d$ nodes to which it connects. A more detailed discussion is provided in Section V.

The rest of the paper is organized as follows. Section II presents the general approach followed in the paper for code construction and for proving information-theoretic secrecy. Section III presents the secure MBR code for all parameters $[n, k, d]$ and $\{\ell, \ell'\}$. Section IV presents the secure MSR codes for all parameters $[n, k, d \geq 2k-2]$ and $\{\ell, \ell'\}$. The paper concludes with a discussion in Section V.

## II. APPROACH

We approach the problem of providing secrecy in the presence of eavesdroppers, in the following manner. To construct a secure code for a given $[n, k, d]$, we choose the corresponding PM code [2] with the same values of system parameters $[n, k, d]$. In the input to the PM code (without secrecy), we replace a specific, carefully chosen set of

$$R = B - B^{(s)} \quad (3)$$

message symbols with $R$ random symbols. Each of these random symbols are chosen uniformly and independently from $\mathbb{F}_q$, and are also independent of the message symbols.

If the random symbols are treated as message symbols, the secure code becomes identical to the original code. Hence, the processes of reconstruction and repair in the secure code can be carried out in the same way as in the original code.

To prove $\{\ell, \ell'\}$ secrecy of our codes, we consider the worst case scenario where an eavesdropper has access to precisely $\{\ell, \ell'\}$ nodes. Let $\mathcal{U}$ denote the collection of the $B^{(s)}$ message symbols, and let $\mathcal{R}$ denote the collection of $R$ random symbols as defined in (3). Further, let $\mathcal{E}$ denote the collection of symbols that the eavesdropper gains access to. For each of the codes presented in this paper, the proof of information-theoretic secrecy proceeds in the following manner. All logarithms are taken to the base $q$.

Step 1: We show that given all the message symbols $\mathcal{U}$ as side-information, the eavesdropper can recover all the $R$ random symbols, i.e., $H(\mathcal{R}|\mathcal{E},\mathcal{U}) = 0$.

Step 2: Next we show that all but $R$ of the symbols obtained by the eavesdropper are functions of these $R$ symbols, i.e., $H(\mathcal{E}) \leq R$.

Step 3: We finally show that the two conditions listed in steps 1 and 2 above necessarily implies that the mutual information between the message symbols $\mathcal{U}$ and the symbols $\mathcal{E}$ obtained by the eavesdropper is zero, i.e., $I(\mathcal{U};\mathcal{E}) = 0$.

## III. SECURE MBR CODES FOR ALL $[n, k, d]$, $\{\ell, \ell'\}$

MBR codes achieve the minimum possible repair-bandwidth: a replacement node downloads only what it stores, i.e., have $d\beta = \alpha$. Substituting this in the bound in (1), and replacing the inequality with equality, we get that in the absence of secrecy requirements an MBR code must satisfy

$$B = \left(kd - \binom{k}{2}\right)\beta, \ \alpha = d\beta . \quad (4)$$

In this section, we present explicit constructions of information-theoretically secure MBR codes for all parameter values $[n, k, d]$ and all $\{\ell, \ell'\}$. These codes meet the upper bound (2) on the total number of message symbols, thus showing that (2) is indeed the secrecy capacity at the MBR point for all parameters. These codes are based on the PM-MBR codes constructed in [2]. We first provide a brief description of the PM-MBR codes, before moving on to the construction of the secure PM-MBR codes.

We construct codes for the case $\beta = 1$, and codes for any higher value of $\beta$ can be obtained by a simple concatenation of the $\beta = 1$ code. In the terminology of distributed storage, this process is known as *striping*. Thus an MBR code with $\beta = 1$ has $\alpha = d$.

### A. Recap of the Product-Matrix MBR codes

The PM-MBR code [2] can be described in terms of an $(n \times \alpha)$ code matrix $C$, where the $\alpha$ elements in its $i^{\text{th}}$ row represent the $\alpha$ symbols stored in node $i$ $(1 \leq i \leq n)$. The code matrix $C$ is a product of two matrices: a fixed $(n \times d)$ encoding matrix $\Psi$ and a $(d \times \alpha)$ message matrix $M$ comprising the $B$ message symbols in a possibly redundant fashion, i.e.,

$$C = \Psi M . \quad (5)$$

Denoting the $i^{\text{th}}$ row of $\Psi$ as $\underline{\psi}_i^t$, the $\alpha$ symbols stored in the $i^{\text{th}}$ storage node is expressed as $\underline{\psi}_i^t M$. The superscript '$t$' denotes the transpose of a matrix.

In the PM-MBR code, the encoding matrix $\Psi$ and the message matrix $M$ are of the form

$$\underbrace{\Psi}_{n \times d} = \left[\ \underbrace{\Phi}_{n \times k}\ \ \underbrace{\Delta}_{n \times (d-k)}\ \right], \ \underbrace{M}_{d \times d} = \left[\begin{array}{cc} \overbrace{S}^{k \times k} & \overbrace{T}^{k \times (d-k)} \\ \underbrace{T^t}_{(d-k) \times k} & \underbrace{0}_{(d-k) \times (d-k)} \end{array}\right]$$

The matrices $\Phi$ and $\Delta$ are chosen in such a way that (a) any $k$ rows of $\Phi$ are linearly independent, and (b) any $d$ rows of $\Psi$ are linearly independent. These requirements can be met, for example, by choosing $\Psi$ to be either a Cauchy or a Vandermonde matrix. The choice of the matrix $\Psi$ governs the choice of the size $q$ of the finite field $\mathbb{F}_q$, e.g., choosing $\Psi$ as Vandermonde allows us to use any $q \geq n$.

The matrices $S$ and $T$ in the message matrix $M$ are populated by the $B$ message symbols,

$$B = kd - \binom{k}{2} = k(d-k) + \frac{k(k+1)}{2} , \quad (6)$$

as follows. The $\frac{k(k+1)}{2}$ symbols in the upper triangular half of the $(k \times k)$ *symmetric* matrix $S$ and the $k(d-k)$ elements in the $(k \times (d-k))$ matrix $T$ are set equal to the $B$ message symbols. Note that the symmetry of matrix $S$ makes $M$ also symmetric.

*Example 1:* We illustrate the code with an example; this example will also be used subsequently to illustrate the secure code. Let $n = 6$, $k = 3$, $d = 4$. Then with $\beta = 1$, we get $\alpha = d = 4$ and $B = 9$. We design the code over the finite field $\mathbb{F}_7$. The $(6 \times 4)$ encoding matrix $\Psi$ can be chosen as a Vandermonde matrix with its $i^{th}$ row as $\underline{\psi}_i^t = [1\ \ i\ \ i^2\ \ i^3]$. The matrices $S$ and $T$, and hence the message matrix $M$ are populated by the 9 message symbols $\{u_i\}_{i=1}^9$ as

$$S = \begin{bmatrix} u_1 & u_2 & u_3 \\ u_2 & u_4 & u_5 \\ u_3 & u_5 & u_6 \end{bmatrix}, T = \begin{bmatrix} u_7 \\ u_8 \\ u_9 \end{bmatrix}, M = \begin{bmatrix} u_1 & u_2 & u_3 & u_7 \\ u_2 & u_4 & u_5 & u_8 \\ u_3 & u_5 & u_6 & u_9 \\ u_7 & u_8 & u_9 & 0 \end{bmatrix}.$$

We now describe the reconstruction and the repair processes in the PM-MBR code.

*1) Reconstruction:* Let $\Psi_{\text{DC}} = \left[\ \Phi_{\text{DC}}\ \ \Delta_{\text{DC}}\ \right]$ be the $(k \times d)$ submatrix of $\Psi$, corresponding to the $k$ rows of $\Psi$ to which the data-collector connects. Thus the data-collector has access to the symbols $\Psi_{\text{DC}} M = \left[\ \Phi_{\text{DC}} S + \Delta_{\text{DC}} T^t\ \ \Phi_{\text{DC}} T\ \right]$. By construction, the matrix $\Phi_{\text{DC}}$ is nonsingular. Hence, by multiplying the matrix $\Psi_{\text{DC}} M$ on the left by $\Phi_{\text{DC}}^{-1}$, one can recover first the matrix $T$ and subsequently, the matrix $S$.

*2) Repair:* Let $\underline{\psi}_f^t$ be the row of $\Psi$ corresponding to the failed node $f$. Thus the $d$ symbols stored in the failed node are $\underline{\psi}_f^t M$. The replacement for the failed node $f$ connects to an arbitrary set $\{h_i | 1 \leq i \leq d\}$ of $d$ remaining nodes. Each of these $d$ nodes passes on the inner product $(\underline{\psi}_{h_i}^t M)\underline{\psi}_f$ to the replacement node. Thus from these $d$ nodes, the replacement node obtains the $d = \alpha$ symbols $\Psi_{\text{rep}} M \underline{\psi}_f$, where $\Psi_{\text{rep}} = \left[\underline{\psi}_{h_1} \cdots \underline{\psi}_{h_d}\right]^t$. By construction, the $(d \times d)$ matrix $\Psi_{\text{rep}}$ is invertible. This allows the replacement node to recover $M\underline{\psi}_f$. Since $M$ is symmetric, $(M\underline{\psi}_f)^t = \underline{\psi}_f^t M$ which is precisely the data stored in the node prior to failure.

### B. Information-theoretic Secrecy in the PM-MBR Code

For the MBR code, we have $d\beta = \alpha$, i.e., a replacement node stores all the data that it downloads during its repair. Thus an eavesdropper does not obtain any extra information from the data that is downloaded for repair. Hence for an MBR code, we can assume without loss of generality that $\ell' = 0$.

In this section, we will construct codes that achieve the upper bound in (2) at the MBR point. Substituting $\alpha = d\beta$

in (2) and replacing the inequality with equality, we get that such a code must necessarily satisfy

$$B^{(s)} = \left(kd - \binom{k}{2}\right)\beta - \left(ld - \binom{\ell}{2}\right)\beta . \quad (7)$$

We now construct an $\{\ell, \ell'\}$ secure MBR code satisfying (7), based on the PM-MBR code. We denote the PM-MBR code [2] described above as $\mathcal{C}$, and the secure PM-MBR code constructed here as $\mathcal{C}^{(s)}$. As mentioned previously, we will present the construction for the case $\beta = 1$.

Let $\Psi^{(s)}$ be the $(n \times d)$ encoding matrix of code $\mathcal{C}^{(s)}$. Choose $\Psi^{(s)}$ to satisfy the following property in addition to those required by $\Psi$: when restricted to the first $\ell$ columns, any $\ell$ rows are linearly independent. The choice of $\Psi^{(s)}$ as a Cauchy or Vandermonde matrix satisfies this additional property as well. We now modify the message matrix $M$ of code $\mathcal{C}$ to obtain message matrix $M^{(s)}$ of code $\mathcal{C}^{(s)}$. Replace the

$$R = B - B^{(s)} = ld - \binom{\ell}{2} \quad (8)$$

message symbols in the first $\ell$ rows (and hence first $\ell$ columns) of the symmetric matrix $M$ by $R$ random symbols. Each random symbol is chosen independently and uniformly across the elements of $\mathbb{F}_q$. Thus the $(n \times \alpha)$ code matrix for the secure PM-MBR code $\mathcal{C}^{(s)}$ is given by $C^{(s)} = \Psi^{(s)} M^{(s)}$.

*Example 2:* We will use the PM-MBR code in Example 1 to obtain a secure PM-MBR code for $[n = 6, k = 3, d = 4]$ with $\ell = 1$. From (7) with $\beta = 1$ we get $B^{(s)} = 5$. Thus we have $R = B - B^{(s)} = 4$. We replace the four message symbols $u_1, u_2, u_3$ and $u_7$ in Example 1 with random symbols $r_1, r_2, r_3$ and $r_7$ drawn uniformly and independently from $\mathbb{F}_7$ to get the new message matrix $M^{(s)}$ as:

$$M^{(s)} = \begin{bmatrix} r_1 & r_2 & r_3 & r_7 \\ r_2 & u_4 & u_5 & u_8 \\ r_3 & u_5 & u_6 & u_9 \\ r_7 & u_8 & u_9 & 0 \end{bmatrix} . \quad (9)$$

Since the matrix $\Psi$ in Example 1 is a Vandermonde matrix which already satisfies the additional property, we retain it in the new code, i.e., $\Psi^{(s)} = \Psi$. Thus the secure PM-MBR code for the desired parameters is given by $C^{(s)} = \Psi^{(s)} M^{(s)}$.

The following theorems prove the properties of reconstruction, repair and secrecy in the secure PM-MBR code.

*Theorem 1 (Reconstruction and Repair):* In code $\mathcal{C}^{(s)}$ presented above, a data-collector can recover all the $B^{(s)}$ message symbols by downloading data stored in any $k$ nodes, and a failed node can be repaired by downloading one symbol each from any $d$ remaining nodes.

*Proof:* Treating the random symbols also as message symbols, the secure PM-MBR code $\mathcal{C}^{(s)}$ becomes identical to the PM-MBR code $\mathcal{C}$. Thus reconstruction and repair in $\mathcal{C}^{(s)}$ are identical to that in $\mathcal{C}$. ∎

*Theorem 2 (Information-theoretic Secrecy):* In code $\mathcal{C}^{(s)}$ designed for a given value of $\ell$, an eavesdropper having access to at most $\ell$ nodes gets no information pertaining to the message.

*Proof:* Let $\Psi_{\text{eve}}^{(s)}$ be the $(\ell \times d)$ submatrix of $\Psi^{(s)}$, corresponding to the $\ell$ rows of $\Psi$ to which the eavesdropper has gained access. Thus the eavesdropper has access to the $\ell d$ symbols in the $(\ell \times d)$ matrix $E^{(s)}$ defined as

$$E^{(s)} = \Psi_{\text{eve}}^{(s)} M^{(s)} . \quad (10)$$

Following the approach described in Section II, we first show that given the message symbols as side information, an eavesdropper can decode all the random symbols. To this end, define $\tilde{M}^{(s)}$ as a $(d \times d)$ matrix obtained by setting all message symbols in $M^{(s)}$ to zero. Thus $\tilde{M}^{(s)}$ has its first $\ell$ rows and first $\ell$ columns identical to that of $M^{(s)}$, and zeros elsewhere. Let

$$\tilde{E}^{(s)} = \Psi_{\text{eve}}^{(s)} \tilde{M}^{(s)} , \quad (11)$$

which are the $\ell d$ symbols that the eavesdropper has access to, given the message symbols as side information. Recall the property of $\Psi_{\text{eve}}^{(s)}$ wherein any $\ell$ rows, when restricted to the first $\ell$ columns, are independent. Thus, recovering the $R$ random symbols from $\tilde{E}$ is identical to data reconstruction in the original PM-MBR code $\hat{\mathcal{C}}$ designed for $[\hat{n} = n, \hat{k} = \ell, \hat{d} = d]$, $\hat{\ell} = 0$. Thus, given the message symbols, the eavesdropper can decode all the random symbols.

The next step is to show that $H(\mathcal{E}) \leq R$. From the value of $R$ in (8), it suffices to show that out of the $\ell d$ symbols that the eavesdropper has access to, $\binom{\ell}{2}$ of them are functions (linear combinations) of the rest. Consider, the $(\ell \times \ell)$ matrix

$$E^{(s)} (\Psi_{\text{eve}}^{(s)})^t = \Psi_{\text{eve}}^{(s)} M^{(s)} (\Psi_{\text{eve}}^{(s)})^t . \quad (12)$$

Since $M^{(s)}$ is symmetric, the $(\ell \times \ell)$ matrix in (12) is also symmetric. Thus $\binom{\ell}{2}$ dependencies among the elements of $E^{(s)}$ can be described by the $\binom{\ell}{2}$ upper-triangular elements of the expression

$$E^{(s)} (\Psi_{\text{eve}}^{(s)})^t - \Psi_{\text{eve}}^{(s)} (E^{(s)})^t = 0 . \quad (13)$$

Using the linear-independence property of the rows of $\Psi^{(s)}$, it can be shown that these $\binom{\ell}{2}$ redundant equations are linearly independent. Thus the eavesdropper has access to at-most $\ell d - \binom{\ell}{2}$ independent symbols, i.e., $H(\mathcal{E}) \leq R$.

We have shown that in the secure PM-MBR code, steps 1 and 2 of the approach described in Section II hold true. The final part of the proof, Step 3, establishes that the eavesdropper obtains no information about the message.

$$\begin{aligned} I(\mathcal{U}; \mathcal{E}) &= H(\mathcal{E}) - H(\mathcal{E}|\mathcal{U}) & (14) \\ &\leq R - H(\mathcal{E}|\mathcal{U}) & (15) \\ &= R - H(\mathcal{E}|\mathcal{U}) + H(\mathcal{E}|\mathcal{U}, \mathcal{R}) & (16) \\ &= R - I(\mathcal{E}; \mathcal{R}|\mathcal{U}) \\ &= R - (H(\mathcal{R}|\mathcal{U}) - H(\mathcal{R}|\mathcal{E}, \mathcal{U})) & (17) \\ &= R - H(\mathcal{R}|\mathcal{U}) & (18) \\ &= R - R & (19) \\ &= 0 , & (20) \end{aligned}$$

where (15) follows from the result of Step 2; (16) follows since every symbol in the system is a function of $\mathcal{U}$ and $\mathcal{R}$, giving $H(\mathcal{E}|\mathcal{U}, \mathcal{R}) = 0$; (18) follows from the result of Step 1; and (19) follows since the random symbols are independent of the message symbols. ∎

## IV. SECURE MSR CODES FOR ALL $[n, k, d \geq 2k-2]$, $\{\ell, \ell'\}$

MSR codes achieve the minimum possible storage at each node. Since a data-collector connecting to any $k$ nodes should be able to recover all the $B$ message symbols, each node must necessarily store at-least a fraction $\frac{1}{k}$ of the entire data. Hence for an MSR code we have $\alpha = \frac{B}{k}$. It follows from (1) (replacing the inequality with equality) that in the absence of secrecy requirements an MSR code must satisfy

$$B = k\alpha, \quad d\beta = \alpha + (k-1)\beta . \quad (21)$$

From (21) we see that, in general, for an MSR code $d\beta > \alpha$. Thus the amount of data downloaded during repair is greater than what is eventually stored. This requires us to distinguish between the situations when the eavesdropper has access to only the data stored in a node, and when it has access to the data downloaded during repair. Note that the data downloaded by a replacement node during repair also contains the data that is eventually stored in it.

In this section we present explicit constructions of information-theoretically secure MSR codes for all parameter values $[n, k, d \geq 2k-2]$ and all $\{\ell, \ell'\}$. The secure MSR codes are based on the PM-MSR codes presented in [2].

### A. Recap of the Product-Matrix MSR codes

We first provide a brief description of the PM-MSR code [2]. The code is designed for the case $d = 2k-2$, and can be extended to $d > 2k-2$ via shortening (see [2], [5] for a detailed description of shortening in MSR codes). As in the MBR case, we construct codes for the case when $\beta = 1$. Setting $d = 2k-2$ and $\beta = 1$ in (21) gives

$$B = \alpha(\alpha+1), \quad \alpha = k-1, \quad d = 2\alpha . \quad (22)$$

The PM-MSR code $\mathcal{C}$ in [2] can be described in terms of an $(n \times \alpha)$ code matrix $C = \Psi M$, with the $i^{th}$ row of $C$ containing the $\alpha$ symbols stored in node $i$. The $(n \times d)$ encoding matrix $\Psi$ is of the form $\Psi = [\Phi \ \Lambda\Phi]$, where $\Phi$ is an $(n \times \alpha)$ matrix and $\Lambda$ is an $(n \times n)$ diagonal matrix satisfying: (a) any $\alpha$ rows of $\Phi$ are linearly independent, (b) any $d$ rows of $\Psi$ are linearly independent, and (c) the diagonal elements of $\Lambda$ are all distinct. The $((d = 2\alpha) \times \alpha)$ message matrix $M$ is of the form $M = [S_1 \ S_2]^t$, where $S_1$ and $S_2$ are $(\alpha \times \alpha)$ symmetric matrices. The two matrices $S_1$ and $S_2$ together contain $\alpha(\alpha+1)$ distinct symbols, and these positions are populated by the $B = \alpha(\alpha+1)$ message symbols. This completes the description of the code construction.

A description of the reconstruction and repair operations under this code can be found in [2]. The repair algorithm in [2] is such that the data downloaded by any node for repair is independent of the set of $d$ nodes to which it connects. This property is highly advantageous while constructing secure codes, as discussed in Section V.

### B. Information-theoretic Secrecy in the PM-MSR Code

For the MSR case, from (2) we get

$$B^{(s)} \leq (k-\ell)\alpha . \quad (23)$$

On the other hand, the $\{\ell, \ell'\}$ secure MSR codes constructed in the present paper (for $d \geq 2k-2$) achieve

$$B^{(s)} = (k-\ell)(\alpha - \ell'\beta) . \quad (24)$$

Thus our codes are optimal for $\ell' = 0$. As mentioned previously, it is unknown at present as to whether or not, our codes are optimal when $\ell' \geq 1$.

The expression for $B^{(s)}$ in (24) can be interpreted as follows. Consider a data-collector attempting to reconstruct the message from the data stored in some $k$ nodes, and an eavesdropper having access to some $\ell$ of these $k$ nodes. These $\ell$ nodes will not provide any useful information, thus resulting in the first term $(k-\ell)$ in the product. Furthermore, the eavesdropper may have access to the data passed for repair of some $\ell'$ of the $\ell$ nodes, and hence to the $\ell'\beta$ (potentially distinct) symbols passed by each of the remaining $(k-\ell)$ nodes during repair. These symbols should not reveal any information, and hence the second term $(\alpha - \ell'\beta)$.

We now describe the construction of the secure PM-MSR code (for $\beta = 1$). We retain the notation used in Section III-B. Choose $\Psi^{(s)}$ such that it satisfies the following property in addition to those required for $\Psi$: when restricted to the first $\ell$ columns, any $\ell$ rows of $\Psi^{(s)}$ are linearly independent. Next, define a collection $\mathcal{R}$ of

$$R = B - B^{(s)} = \ell\alpha + (k-\ell)\ell' \quad (25)$$

random symbols picked independently with a uniform distribution over the elements of $\mathbb{F}_q$, where (25) follows from (21) and (24). Use these $R$ random symbols to replace the following $R$ symbols in the message matrix $M$ of code $\mathcal{C}$, to obtain matrix $M^{(s)}$: the $\ell\alpha - \binom{\ell}{2}$ symbols in the first $\ell$ rows (and hence the first $\ell$ columns) of the symmetric matrix $S_1$, the $\binom{\ell}{2}$ symbols in the intersection of the first $(\ell-1)$ rows and first $(\ell-1)$ columns of the symmetric matrix $S_2$, and the $(k-\ell)\ell'$ remaining symbols in the first $\ell'$ rows (and hence the first $\ell'$ columns) of $S_2$. The secure PM-MSR code is given by $C^{(s)} = \Psi^{(s)}M^{(s)}$.

The following theorems prove the properties of reconstruction, repair and secrecy in the secure PM-MSR code.

*Theorem 3 (Reconstruction and Repair):* In code $\mathcal{C}^{(s)}$ presented above, a data-collector can recover all the $B^{(s)}$ message symbols by downloading data stored in any $k$ nodes, and a failed node can be repaired by downloading one symbol each from any $d$ remaining nodes.

*Proof:* As in the proof of Theorem 1, treating the random symbols also as message symbols, the secure PM-MSR code $\mathcal{C}^{(s)}$ becomes identical to the PM-MSR code $\mathcal{C}$. Thus reconstruction and repair in $\mathcal{C}^{(s)}$ are identical to that in $\mathcal{C}$. ∎

*Theorem 4 (Information-theoretic Secrecy):* In code $\mathcal{C}^{(s)}$ designed for a given value of $\ell$, an eavesdropper having access to at most $\ell$ nodes gets no information pertaining to the message.

*Proof (Sketch):* Let $\Psi^{(s)}_{\text{eve}}$ be the $(\ell \times d)$ submatrix of $\Psi^{(s)}$, corresponding to the $\ell$ rows of $\Psi$ to which the eavesdropper has gained access. Further, let $\Phi^{(s)}_{\text{eve1}}$ be the $(\ell' \times \alpha)$ submatrix of $\Phi^{(s)}$, corresponding to the $\ell'$ nodes in which the eavesdropper has access to the repair downloads as well. Note that by definition of an $\{\ell, \ell'\}$ secure system, these $\ell'$ nodes are a subset of the set of $\ell$ nodes that constitute the matrix $\Psi_{\text{eve}}^{(s)}$. From the repair algorithm of the PM-MSR code of [2], it turns out that the symbols $\mathcal{E}$ that the eavesdropper gains access to comprises the elements of the $(\ell \times \alpha)$ matrix $\Psi_{\text{eve}}^{(s)} M$ and the elements of the $(d \times \ell')$ matrix $M(\Phi_{\text{eve1}}^{(s)})^t$.

Following the approach described in Section II, and in a manner analogous to the proof of Theorem 2, it can first be shown that given the message symbols as side information, an eavesdropper can decode all the random symbols. Next, using the properties of the matrix $\Psi^{(s)}$ and the specific structure of the message matrix $M^{(s)}$, it can also be shown that $H(\mathcal{E}) \leq R$. Finally, the arguments in (14) to (20) established that the eavesdropper obtains no information about the message. ∎

The extension to the case $d > 2k - 2$ can be achieved via shortening ([2], [5]), using which one can use any linear secure MSR code with parameters $[n+1, k+1, d+1, \ell+1, \ell']$ to construct a linear secure MSR code for parameters $[n, k, d, \ell, \ell']$.

## V. DISCUSSION

The Product-Matrix framework [2] possesses two particular attributes that make the codes built in this framework attractive from the security perspective. First, many codes in the literature including those in [1] consider *functional repair*, wherein the data stored in the replacement node is permitted to be different from that of the failed node as long as it satisfies the reconstruction and functional-repair properties of the system. This allows an eavesdropper to gain a greater amount of information by reading the data stored in a node across multiple instances of repair. On the other hand, PM codes offer *exact-repair*, wherein the data stored in the replacement node is identical to that in the failed node. Second, even if repair is exact, the data downloaded during repair of a particular node may depend on the set of $d$ nodes helping in the repair process, and hence may be different during different instances of repair of that node. The PM framework, by design, ensures that the information contained in the symbols downloaded by the replacement node is independent of the identities of the helper nodes. This restricts information exposed to an eavesdropper that has access to the data downloaded during repair.